\documentclass[12pt]{article}
\usepackage[doublespacing]{setspace}
\usepackage{graphicx, amsmath, subfigure}

\begin{document}

\title{Landscape construction in non-gradient dynamics: A model from evolution\thanks{Running head: Landscape in non-gradient dynamics} \thanks{The authors declare that there is no conflict of interest.} \thanks{
We give special thanks to Dr. Ping Ao for his constructive comments and suggestions.
We also thank the anonymous referee for helpful suggestions on the presentation of the work.
This work is partly supported by the China Scholarship Council (File No. 201408020004).}
}

\author{Song Xu\thanks{
Corresponding author. Email: sxu11@ucla.edu.} 
\thanks{Department of Biomathematics, University of California at Los Angeles, Los Angeles, CA 90095, USA
}, Xinan Wang\thanks{
Computer Science and Engineering Department, University of California San Diego, La Jolla, CA 92093, USA
}, Shuyun Jiao\thanks{
Shanghai Center for Systems Biomedicine, Key Laboratory of Systems Biomedicine of Ministry of Education, Shanghai Jiao Tong University, 200240, Shanghai, China}
\thanks{
Department of Mathematics, Xinyang Normal University, 464000, Xinyang, Henan, China.
}}

\maketitle

\newpage

\begin{abstract}
Adaptive landscape has been a fundamental concept in many branches of modern biology since Wright's first proposition in 1932. Meanwhile, the general existence of landscape remains controversial. 
The causes include the mixed uses of different landscape definitions with their own different aims and advantages. Sometimes the difficulty and the impossibility of the landscape construction for complex models are also equated. 
To clarify these confusions, based on a recent formulation of Wright's theory, the current authors construct generalized adaptive landscape in a two-loci population model with non-gradient dynamics, where the conventional gradient landscape does not exist. 
On the generalized landscape, a population moves along an evolutionary trajectory which always increases or conserves adaptiveness but does not necessarily follow the steepest gradient direction. 
Comparisons of different aspects of various landscapes lead to a conclusion that the generalized landscape is a possible direction to continue the exploration of Wright's theory for complex dynamics.
\end{abstract}

\section{Introduction}
Wright (1932) proposed adaptive landscape to illustrate his shifting-balance theory of evolution. It maps a measure of the population adaptiveness as the ``height" of the landscape over the allele or genotypic frequency space and visualizes a specific population as a dot on the landscape (Arnold et al., 2001). A peak on the landscape represents a local adaptive state and a valley represents an unadaptive state. Evolution is generally visualized as a valley-to-peak uphill climbing process.
Biologists extensively used landscape to visualize and quantify stable states and evolutionary trajectories in the fields of population biology (Lande, 1976; Ao, 2005; Weinreich et al., 2006), developmental biology (Waddington, 1957), gene regulation (Delbr{\'u}ck, 1949; Zhu et al., 2004), neural dynamics (Hopfield, 1999), protein folding (Dill and Chan, 1997), cell cycle (Wang et al., 2006), signal transduction pathway (Lapidus et al., 2008), and cancer genesis and progression (Andrews, 2002; Ao et al., 2008).
It is found to be analogous to the potential energy function, which characterizes the long-term stability of a physical systems (Ao, 2008; Wang et al., 2008; Ge and Qian, 2012). It also connects to the Lyapunov function in control theory (Lyapunov 1992; Haddad and Chellaboina, 2008), which is a measure for the local stability of a fixed point in dynamical systems.

Although it has been widely employed for similar visualization and quantification purposes in different contexts, the adaptive landscape was not always unambiguously defined and consistently used (Rice, 2004). The interchanging uses of different landscape definitions, implicitly or explicitly, may lead to confusions and controversies. Here we summarize different types of population-level landscapes as follows: 

The mean fitness landscape (the``first type") maps the effect of natural selection over the frequency space, where the adaptive peaks are also called ``fitness peaks", denoting the environmentally favorable population states. This definition is most intuitive in terms of biological meaning, and is very useful when there is only selection or selection is very strong. 

Another type of landscape is also defined on the population level. Compared to the fitness landscape, it does not necessarily map a population's mean fitness, but maps a more general measure of a population's evolutionary stability (the ``second type"). Such measure usually integrates the effects of all biological forces (selection, mutation, and recombination) instead of being limited to selection. 
It visualizes evolution as uphill climbing process (called the Lyapunov property) and might be termed ``generalized landscape".
Ao (2004), Qian (2005), Wang et al. (2008), and Barton and Coe (2009) took efforts to construct such Lyapunov- or energy- like landscape in biological systems. Ao (2004) first explicitly revealed the dynamical structure built into an evolution model, where a generalized landscape was constructed as an intrinsic component. Many fruitful results were obtained following his work in both linear-dynamical systems (Kwon et al., 2005) and systems with non-linear dynamics (Tang et al., 2013; Ma et al., 2014). It was applied to several biological systems (Zhu et al., 2004; Ao, 2005; Ao et al., 2008; Xu et al., 2014) and was validated in a physical experiment (Volpe et al., 2010).

In simple models with gradient dynamics, a generalized landscape reduces to a ``gradient landscape" (the ``third type"), on which a population's trajectory not only follows the uphill direction, but is strictly the steepest uphill (gradient) direction of the landscape. However, such gradient landscape only exists in gradient models, which represent a very small subset of all biological models. The term ``gradient" here refers to the multi-dimensional derivative of a scalar measure function. It should be distinguished from the term ``selection gradient" in population biology, which usually denotes the steepness of the individual-level fitness landscape (Kirkpatrick and Rousset, 2005).

The differences in the definitions and constraints of the above three landscapes sometimes cause confusion. For example, Moran (1964) derived the mean fitness landscape and expected it to act as a generalized landscape. Many researchers implicitly used the mean fitness landscape as a gradient landscape (Edwards, 2000). Gavrilets (2004) and Carneiro and Hartl (2010) felt that generalized landscape is very useful but is hard to construct. Kaplan (2008) doubted whether we should use this concept. Recently, Weinreich et al. (2013) challenged the validity of adaptive landscape again in a 2-loci population model by showing that the model is not gradient.

Here we do not aim at providing a universal guide for finding adaptive landscape for every non-gradient system or exhausting all its biological applications in one paper, but trying to explore the applicability of adaptive landscape in non-gradient dynamics. We take the same population model as in Weinreich et al. (2013)'s work and construct generalized landscapes. We also clarify and compare the differences of various landscapes to avoid confusions. The current work is the first one to our knowledge to voice this out in a specific biology model. 

\section{Population genetics model}
\label{sec:model_main}

The model describes a population's evolution on two loci:
the $A/a$ pair and the $B/b$ pair. 
It was used to examine the dynamical system describing an infinite population’s evolutionary trajectory in response to natural selection and recombination for the two-loci genotypic fitness landscape in different coordinates (Weinreich et al., 2013).
It assumes continuous time and non-overlapping generations for convenient analysis of the dynamics of the model.
The population size is assumed to be infinite, so there is no random effect induced by genetic drift. There is no mutation considered in the system. Under various biological forces such as selection and recombination, the frequencies of different genotypes $p_{AB},~p_{Ab},~p_{aB},$ and $p_{ab}$ may change with time. Denote $p_1 = p_{AB}(t),~ p_2 = p_{Ab}(t),~ p_3 = p_{aB}(t),~ p_4 = p_{ab}(t)$, where the time-dependency is neglected without ambiguity. Let $w_i$ denote the fitness of each genotype ($i=1,~2,~3,~4$). Under selection alone, the dynamics of the four-dimension system is
\begin{equation}
\label{Eq:4d}
\dot{p}_i=\dfrac{w_i\cdot p_i}{\overline{W}}-p_i, \qquad \qquad i=1,\dots,4.
\end{equation}
where $\dot p_i$ is the time-derivative of $p_i$. $\overline{W}$ is the population mean fitness
\begin{equation}
\label{Eq:W4d}
\overline{W}=\sum_{i=1}^4w_i\cdot p_i.
\end{equation}
In this simple case, $\overline{W}$ is a valid adaptive landscape (Moran 1964). If we further add recombination, the system in Eq.~(\ref{Eq:4d}) becomes
\begin{eqnarray}
\dot{p}_1=\dfrac{w_1\cdot p_1}{\overline{W}} - p_1 - r(p_1\cdot p_4-p_2\cdot p_3), \\
\dot{p}_2=\dfrac{w_2\cdot p_2}{\overline{W}} - p_2 + r(p_1\cdot p_4-p_2\cdot p_3), \\
\dot{p}_3=\dfrac{w_3\cdot p_3}{\overline{W}} - p_3 + r(p_1\cdot p_4-p_2\cdot p_3), \\
\dot{p}_4=\dfrac{w_4\cdot p_4}{\overline{W}} - p_4 - r(p_1\cdot p_4-p_2\cdot p_3).
\end{eqnarray}
where $r$ is the recombination rate satisfying $0\leq r\leq 0.5$. Recombination induces complex interactions between different loci, leading to the difficulty for landscape construction (Moran 1964). The four variables $p_i$ are constrained by the probability conservation
\begin{equation}
p_1 + p_2 + p_3 + p_4 = 1. \label{Eq:constr1}
\end{equation}
If one further assumes symmetric selections on $Ab$ and $aB$ individuals ($w_2=w_3$) and symmetric initial frequency distribution between $Ab$ and $aB$ ($p_2(0)=p_3(0)$), we have the symmetricity holds for any $t>0$:
\begin{equation}
p_2(t) = p_3(t). \label{Eq:constr2}
\end{equation}
With the two constraints in Eq.~(\ref{Eq:constr1}) and Eq.~(\ref{Eq:constr2}), we are able to express the model in a two-dimensional space:
\begin{eqnarray}
&& \dot p_1 = \frac{w_1p_1}{\overline W}-p_1-rp_1+r(p_1+p_2)^2, \label{Eq:origMod1} \\
&& \dot p_2 = \frac{w_2p_2}{\overline W}-p_2+rp_1-r(p_1+p_2)^2. \label{Eq:origMod2}
\end{eqnarray}
Equivalently, it is possible to transform the two-dimensional system in Eq.~(\ref{Eq:origMod1}) and Eq.~(\ref{Eq:origMod2}) into a different coordinate space (Weinreich et al., 2013):
\begin{eqnarray}
&& q_1=p_1+p_2, \label{Eq:transMod1} \\
&& q_2=p_1\cdot p_4-p_2\cdot p_3. \label{Eq:transMod2}
\end{eqnarray}
where $q_1$ takes the meaning of the allele frequency of $A$ and $q_2$ is the linkage-disequilibrium term. 
Such operation is commonly employed to examine what space would more accurately represent the evolutionary trajectories of a model. 
Here we use $q_1$ and $q_2$ to avoid ambiguous notations with other variables. To make sure that $0\leq p_i\leq 1$ for $i=1,2,3,4$, the two variables $q_1$ and $q_2$ in the $q_1\times q_2$ space is confined by 
\begin{eqnarray}
&&0\leq q_1 \label{Eq:q1region} \leq 1, \\
&&\textrm{Max}\{-q_1^2,-(1-q_1)^2\}\leq q_2 \label{Eq:q2region} \leq q_1(1-q_1).
\end{eqnarray}
This gives a sector-shaped domain of $q_1$ and $q_2$ for the transformed system described by Eq.~(\ref{Eq:transMod1}) and Eq.~(\ref{Eq:transMod2}). Figure 1 shows the generalized landscape (grayscale), the gradient flow of landscape (arrows), and a specific evolutionary trajectory (the bold curve) in this domain as an example.

To further specify the model, we follow Weinreich et al. (2013)'s scheme of fitness values as: $w_1=1+s_2$, $w_2=w_3=1+s_1$, and $w_4=1$. Here $s_1$ denotes the selective advantages of the genotypes $Ab$ and $aB$ over the genotype $ab$ (taken to be the same) and $s_2$ is that of the genotype $AB$ over $ab$. The parameter range for $s_1,~s_2$ is taken such that $w_1>w_2=w_3>0$:
\begin{equation}
\label{Eq:param region}
-1<s_1<s_2, \qquad 0<s_2.
\end{equation}
Inside the meaningful parameter ranges, there might exist one or two adaptive peaks in the model, at which we will come back in Section \ref{sec:construction}. The dynamical equations of the transformed system are:
\begin{eqnarray}
&& \dot q_1 = \frac{q_2(s_2-s_1-q_1(s_2-2s_1))+(1-q_1)q_1(q_1(s_2-2s_1)+s_1)}{1+2q_1s_1+(q_2+q_1^2)(s_2-2s_1)}, \label{Eq:q1} \\
&& \dot q_2 = \frac{(1-r)\cdot \bigg(q_2(1+s_2)-(2q_1q_2(1-q_1)-q_2^2-q_1^2(1-q_1)^2)(s_2-2s_1-s_1^2)\bigg)}{\left(1+2q_1s_1+(q_2+q_1^2)(s_2-2s_1)\right)^2} -q_2. \nonumber \\
\label{Eq:q2}
\end{eqnarray}
We denote $\dot{q}_1=f_1,~~\dot{q}_2=f_2$ where $[f_1,f_2]^\tau=\textbf f$ is the deterministic dynamics of the system. We also denote $\textbf q=[q_1,q_2]^\tau$ and $\dot{\textbf q}=[\dot q_1, \dot q_2]^\tau$. 

\section{Generalized adaptive landscape}
\label{sec:framework_main}

\subsection{Decomposition-based construction}
In a deterministic model, the evolution equation at state $\textbf{q}$ is
\begin{equation}
\dot{\textbf{q}}= \textbf{f}(\textbf{q}),
\label{Eq:gener}
\end{equation}
where the notations are consistent with those in the current model, but are actually more general in $n\geq 2$ dimensions. Ao (2004) proposed an explicit decomposition of $\textbf{f}$ into three dynamical components
\begin{equation}
\textbf{f}(\textbf{q})=(D(\textbf{q})+Q(\textbf{q}))\cdot \nabla\phi(\textbf{q}),
\label{Eq:decomp}
\end{equation}
where $\phi$ is a generalized adaptive landscape,
$D$ is a symmetric and semi-positive definite matrix, and $Q$ is an anti-symmetric matrix.

With the decomposition, the dynamics along a specific evolutionary trajectory $\textbf q(t)$ can be separated into two parts:
The dissipative part $\textbf{f}_d(\textbf{q})=D(\textbf{q})\nabla\phi(\textbf{q})$ contributes to the increase of adaptiveness, which is ensured by the semi-definity of $D(\textbf q)$:
\begin{equation}
\dot{\phi_d}(\textbf{q}) = (\nabla\phi(\textbf{q}))^{\tau} \textbf{f}_{d}(\textbf{q})=(\nabla\phi(\textbf{q}))^{\tau} D(\textbf{q})\nabla\phi (\textbf{q}) \geq 0.
\end{equation}
The conservative part $\textbf{f}_c(\textbf{q})=Q(\textbf{q})\nabla\phi(\textbf{q})$ does not change the adaptiveness, which is ensured by the anti-symmetry of $Q(\textbf q)$:
\begin{equation}
\dot{\phi_c}(\textbf{q}) = (\nabla\phi(\textbf{q}))^{\tau} \textbf{f}_{c}(\textbf{q})=(\nabla\phi(\textbf{q}))^{\tau}Q(\textbf{q})\nabla\phi(\textbf{q}) = 0. \label{Eq:phi_c}
\end{equation}
Together we have
\begin{equation}
\dot{\phi}(\textbf{q}) =\left(\nabla\phi(\textbf{q})\right)^\tau\cdot(\textbf{f}(\textbf{q})) = \dot{\phi_d}(\textbf{q}) + \dot{\phi_c}(\textbf{q}) = \dot{\phi_d}(\textbf{q}) \geq 0.
\label{Eq:crit}
\end{equation}
The result $\dot \phi(\textbf{q}(t)) \geq 0$ shows the Lyapunov property of the generalized landscape, meaning that a population always heads for higher adaptive peak (when $\dot{\phi}(\textbf{q}(t))>0$)
or stays on equi-adaptiveness contour (when $\dot{\phi}(\textbf{q}(t))=0$). This manifests Wright's idea that a population's evolution can be visualized as the ``adaptive" movements on a landscape.

To obtain the decompositions for a specific model, one way is to first find $\phi$, after which the two dynamical factors
$D$ and $Q$ can be obtained as
\begin{align}
\label{Eq:D}
D &= \frac{\textbf{f}\cdot \nabla \phi}{\nabla \phi\cdot \nabla \phi} I,\\
\label{Eq:Q}
Q &= \frac{\textbf{f}\times \nabla \phi}{\nabla \phi\cdot \nabla \phi}.
\end{align}
where $I$ is an $n\times n$ identity matrix. The cross product of the two vectors $\textbf{x},\textbf{y} \in R^n$ is
an anti-symmetric matrix in $R^{n\times n}$
defined as $(\textbf{x}\times \textbf{y})_{i,j} = (x_iy_j-x_jy_i)$. Yuan et al. (2013) gave the validity of the constructions.

\subsection{Coordinate-invariance of the Lyapunov property}
\label{Sec:CordInvar}

We found that the Lyapunov property of the generalized landscape is preserved under the coordinate transformation. Suppose that the deterministic system
Eq.~(\ref{Eq:gener}) can be mapped to another dynamical system by an invertible and differentiable coordinate transformation at state $\textbf{q}$:
\begin{equation}
\textbf{y}=\textbf h(\textbf{q}),
\label{Eq:trans}
\end{equation}
Suppose that $\textbf{q}(t)$ is a trajectory in the original system, $\phi(\textbf{q})$ is the generalized landscape for the original system, and $\textbf{y}(t)=\textbf h(\textbf{q}(t))$ is the corresponding trajectory in the new system.
We show that $\psi(\textbf{y})$ defined as
\begin{equation}
\psi(\textbf{y}):=\phi(\textbf h^{-1}(\textbf{y})).
\label{Eq:newLand}
\end{equation}
is a legitimate generalized landscape for the new system $\textbf{y}(t)$. Pick any trajectory $\textbf{y}(t)$ and
two positions $\textbf{y}$ and $\textbf{y}'$ at two different times
$t>t'$. Denote $\textbf{q}=\textbf h^{-1}(\textbf{y})$ and $\textbf{q}'=\textbf h^{-1}(\textbf{y}')$.
The Lyapunov property of $\phi$ for the original system implies that:
\begin{equation}
\phi(\textbf{q})\geq\phi(\textbf{q}')\Rightarrow \phi(\textbf h^{-1}(\textbf{y}))\geq\phi(\textbf h^{-1}(\textbf{y}')) \Rightarrow \psi(\textbf{y})\geq \psi(\textbf{y}'),
\end{equation}
which shows that $\psi$ is a legitimate landscape for the new system $\textbf y(t)$. An example from the current model is Eq.~(\ref{Eq:W-origin}) and Eq.~(\ref{Eq:W}).

\subsection{Coordinate-dependency of the gradient condition}
\label{sec:CordDepend}
Following the notations in Section \ref{Sec:CordInvar}, we prove the coordinate-dependency of the gradient condition. If the original system $\dot{\textbf{q}} = \textbf f(\textbf{q})$ is gradient, the curl of the system should be 0:
\begin{equation}
\nabla \times \textbf f = \partial_1 f_2 - \partial_2 f_1 = 0.
\end{equation}
where $\partial_2 f_1$ denotes the partial derivative of $f_1$ with respect to $q_2$ (similar for $\partial_1 f_2$).
We can denote $\partial_2 f_1 = \partial_1 f_2 = \textbf g(\textbf{q})$.
For the transformed system $\textbf{y}=\textbf h(\textbf{q})$, we have
\begin{equation}
\nabla \times \dot{\textbf{y}} = \nabla \times \dot{\textbf h}(\textbf{q})= \nabla \times \textbf h(\dot{\textbf{q}}) = \nabla \times \textbf h(\textbf f(\textbf{q})).
\label{Eq:tranSysCurl}
\end{equation}
if we assume the function $\textbf h$ is time-independent. Eq.~(\ref{Eq:tranSysCurl}) is not necessarily equal to 0, especially when the transformation $\textbf h$ is nonlinear (as in Eq.~(\ref{Eq:transMod1}) and Eq.~(\ref{Eq:transMod2})). A simple example is to let $ \textbf h(\textbf f(\textbf{q})) = (f_1,~f_2^2)$, then
\begin{equation}
\nabla \times \dot{\textbf{y}} = \partial_1 f_2^2 - \partial_2 f_1 = 2f_2 \partial_1 f_2 - \partial_2 f_1 = (2f_2-1)\textbf g(\textbf{q}).
\end{equation}
which is not necessarily 0. It proves that nonlinear transformation can lead to the failure of the gradient condition.

\section{Landscape construction}
\label{sec:construction}

We construct generalized landscapes in three typical parameter regions, including biologically and mathematically important cases. In the first case (Case 1) with only selection ($r=0,~s_2>s_1>-1$), the mean fitness function $\overline W$ is always a proper generalized landscape for the model.
The second case (Case 2, $r\neq 0,~0\leq s_1<s_2$) with recombination is a single-peak model where we find a direct construction of the landscape, which captures the global stability of the state (1,~0) corresponding to the fixation of the $AB$ genotype. In the third case (Case 3, $r\neq 0,~-1\leq s_1<0<s_2$) with bi-stable selection and recombination, we numerically construct a generalized landscape which visualizes and quantifies the attractive basins of the two peaks and the near-linear boundary between them.

\subsection{Case 1: Selection only ($r=0$)}
\label{Sec:Case-1}

In this case, only selection drives the population to evolve. The mean fitness $\overline{W}$ in Eq.~(\ref{Eq:W4d}) has the Lyapunov property
\begin{eqnarray}
\dot{\overline{W}}&&= \sum_i\dot{p}_i\cdot \dfrac{\partial\overline{W}}{\partial p_i} \nonumber\\
&& = \sum_i\dfrac{p_i}{\overline{W}}\cdot \left(\dfrac{\partial \overline{W}}{\partial p_i}\right)^2-\sum_ip_i\dfrac{\partial \overline{W}}{\partial p_i} \nonumber \\
&& = \dfrac{1}{\overline{W}}\bigg(\sum_ip_i\cdot \left(\dfrac{\partial \overline{W}}{\partial p_i}\right)^2\sum_jp_j-\sum_ip_i\dfrac{\partial \overline{W}}{\partial p_i}\cdot \sum_jp_j\dfrac{\partial \overline{W}}{\partial p_j}\bigg) \nonumber \\
&& = \dfrac{1}{\overline{W}}\bigg(\sum_{i<j}p_ip_j\left(\dfrac{\partial \overline{W}}{\partial p_i}
-\dfrac{\partial \overline{W}}{\partial p_j}\right)^2\bigg)\geq 0.
\label{Eq:W-origin}
\end{eqnarray}
In this case,
we construct the generalized adaptive landscape for the non-gradient model in Eq.~(\ref{Eq:q1}) and Eq.~(\ref{Eq:q2}) as
\begin{equation}
\phi=\overline{W} = 1+2q_1s_1+(q_2+q_1^2)(s_2-2s_1). \label{Eq:W}
\end{equation}
The two dynamical factors are:
\begin{eqnarray}
D = \left(
\begin{array}{cc}
D_{11} & 0 \\
0 & D_{11}
\end{array}
\right)
, \qquad
Q = \left(
\begin{array}{cc}
0 & -Q_{21} \\
Q_{21} & 0
\end{array}
\right).
\end{eqnarray}
according to Eq.~(\ref{Eq:D}) and Eq.~(\ref{Eq:Q}).
We list the analytical expressions for $D_{11}$ and $Q_{21}$ in Appendix \ref{Sec:GQ}. For $s_1=0.08$, we plot the generalized landscape $\phi$ in Figure 2. We also plot the vector field $\textbf{f}$ of the system, its dissipative part $\textbf{f}_d$, and its conservative part $\textbf{f}_c$ in Figure 7. For $s_1=-0.01$, we plot Figure 3 and Figure 8. 

Figure 2 and Figure 3 illustrate that the model has one global stable state at $(1,~0)$. However, it is not biologically obvious because $s_1$ can be negative which may generate an adaptive valley within $q\in (0,1)$ and makes 0 and 1 both adaptive peaks. 
We can rigorously validate this by studying the partial derivatives of the landscape function 
$ \dfrac{\partial \phi}{\partial q_1} = 2s_1+2q_1\cdot(s_2-2s_1), ~
\dfrac{\partial \phi}{\partial q_2} = s_2-2s_1$ in all ranges. 
When $s_2>2s_1\geq0,~s_2>0$, the landscape takes its maximal value on the upper boundary $\{\textbf{q}|~q_2=q_1(1-q_1)\}$. And the landscape $\phi=1+2q_1s_1+(q_1(1-q_1)+q_1^2)\cdot(s_2-2s_1)=1+q_1s_2$
takes its maximal at $(1,0)$. 
When $s_2=2s_1$, the landscape $\phi=1+q_1s_2$ takes its maximal at $(1,0)$.
When $0<s_1<s_2<2s_1$, the landscape takes its maximal value on the lower boundary
$q_2=\textrm{Max}(-q_1^2,-(1-q_1)^2)$ and the global maximal at $(1,0)$.
When $-1<s_1<0$, The landscape $\phi$ takes its maximal on the upper boundary
$q_2=q_1(1-q_1)$, because $s_2>0>s_1$. On the upper boundary,
$\phi=1+q_1s_2$ has only one peak at $(1,0)$.
In conclusion, the landscape $\phi$ has a global adaptive peak in this case.

\subsection{Case 2: One-peak selection with recombination ($s_2>s_1>0,~r>0$)}

Biologically, the fitness scheme: $AB>Ab=aB>ab$ intuitively results in an increasing tendency of the $A$ and $B$ frequencies. With the effect of recombination, we expect this tendency still holds. We can construct the generalized landscape as
\begin{equation}
\phi=q_1,
\label{Fig:lsq1}
\end{equation}
which increases with $q_1$ and is independent of the variable $q_2$ and any other parameters. 
In fact, $\phi=q_1$ is always
non-decreasing inside the region in Eq.~(\ref{Eq:q1region}) and Eq.~(\ref{Eq:q2region}):
\begin{eqnarray}
&& \dot{\phi} = \nabla\phi\cdot \textbf{f} = \frac{\partial\phi}{\partial q_1}f_1 + \frac{\partial\phi}{\partial q_2}f_2 \nonumber \\
&& = f_1 = \frac{q_2(s_2-s_1-q_1(s_2-2s_1))+(1-q_1)q_1(q_1(s_2-2s_1)+s_1)}{1+2q_1s_1+(q_2+q_1^2)(s_2-2s_1)} \nonumber \\
&& = \frac{(q_2+q_1^2)(s_2-s_1)(1-q_1) + s_1q_1[(1-q_1)^2+q_2]}{1+2q_1s_1 - s_1(q_2+q_1^2)+(q_2+q_1^2)(s_2-s_1)} \geq 0 
\label{Eq:single}
\end{eqnarray}
The last inequality comes from the following observations:
The first term of the nominator $(q_2+q_1^2)(s_2-s_1)(1-q_1)$ is positive as $q_2>-q_1^2,~q_1<1,~s_2>s_1$. The second term is positive as $q_2>-(1-q_1)^2, s_1>0, q_1>0$. So the nominator is always bigger or equal to 0. 
For the denominator, using $q_2<q_1(1-q_1)$, we have $- s_1(q_2+q_1^2)>-s_1(q_1(1-q_1)+q_1^2)=-s_1q_1$, so the denominator ends up always bigger than 0 as $1+2q_1s_1-s_1q_1+(q_2+q_1^2)(s_2-s_1)>1+s_1q_1>0$.

The analytical expressions of $D$ and $Q$ are obtained as:
\begin{eqnarray}
D = \left(
\begin{array}{cc}
f_1 & 0 \\
0 & f_1
\end{array}
\right)
, \qquad
Q = \left(
\begin{array}{cc}
0 & -f_2 \\
f_2 & 0
\end{array}
\right).
\end{eqnarray}
according to Eq.~(\ref{Eq:D}) and Eq.~(\ref{Eq:Q}), where $f_1=\dot q_1$ given by Eq.~(\ref{Eq:q1}) and $f_2=\dot q_2$ given by Eq.~(\ref{Eq:q2}).
We plot the generalized landscape $\phi$ in Figure 4 with recombination $r=0.2$. We also plot the vector field $\textbf{f}$ of the system, its dissipative part $\textbf{f}_d$, and its conservative part $\textbf{f}_c$ in Figure 9.

This construction certainly applies to the case with zero recombination rate $r=0,~s_2>s_1>0$ (Figure 10) as a joint case of Case 1 and Case 2, which shows the flexibility of choosing a proper landscape (see Section \ref{sec:discussion} for detailed discussions). The single peak structure captures the global stability of the population which naturally incorporates the local stability analysis by Weinreich et al. (2013). The adaptiveness keeps increasing ($\dot{\phi}>0$), except for the three fixed points $A=(0,0)$, $B=(1,0)$, and $C=(0.5,-0.25)$ where $\dot{\phi}=0$. A is a fixed point corresponding to the fixation of $ab$ and B is a fixed point corresponding to the fixation of $AB$. C is a fixed point only if $r=0$, corresponding to the extreme case where genotypes $p_{AB}=p_{ab}=0,~p_{Ab}=p_{aB}=0.5$. On the landscape, any positive perturbation ($\Delta q_1>0$) will be amplified when the population starts from either A or C, which indicates that these points are unstable. 

\subsection{Case 3: Two-peak selection with recombination $-1<s_1<0<s_2,~r>0$}
\label{Sec:case-3}
Finding a proper landscape under this setting with multiple stable states is more challenging.
We employ a numeric method defined in Eq.~(\ref{Eq:g-distance})
and calculate other dynamical components numerically as well.

In the case with $s_1=-0.01,~s_2=0.1,~r=0.125$, the states $E=(0,0)$ and $F=(1,0)$ are stable fixed states;
the point $G$ between $E$ and $F$ is a saddle point.
The stable manifold of $G$ (the set of points that finally evolve to state $G$) separates the space into the attractive basin of $E$ and the attractive basin of $F$.
In each basin, the geometry length of the orbit from the point $\textbf{q}=\textbf{q}_0$ to the
corresponding stable fixed point is given by
\begin{equation}
\label{Eq:g-distance}
L(\textbf{q}_0)=\int^{+\infty}_0\|\textbf{f}(\tilde{\textbf q}(t,\textbf q_0))\|~dt.
\end{equation}
Here we denote $\tilde{\textbf q}(t,\textbf q_0)$ to emphasize that the specific trajectory starts from the current state $\textbf{q}=\textbf q_0$.

We then assign the landscape adaptiveness (relative ``heights" visualized on the landscape) of each state $\textbf{q}$ along the trajectory proportional to its distance to the stable fixed point:
\begin{equation}
\label{Eq:geo construction}
\phi(\textbf{q}_0):=
\left\{
\begin{array}{rl}
-(L(\textbf{q}_0)-L_E), & \qquad \textrm{If}~~ \lim_{t\to+\infty}\textbf{q}(t)=E,\\
-(L(\textbf{q}_0)-L_F), & \qquad \textrm{Otherwise}.
\end{array} \right.
\end{equation}
$L_E$ and $L_F$ give the adaptiveness values of the stable states $E$ and $F$ (Appendix \ref{BiNumr}). The Lyapunov property is obvious as the adaptiveness always decrease along the trajectory. Actually, after any given time $\delta t>0$ on a specific trajectory:
\begin{eqnarray} 
&&\phi(\textbf q_{\delta t}) - \phi(\textbf q_0) = - (L(\textbf q_{\delta t}) - L(\textbf q_0)) \nonumber \\
&& = - (\int^{+\infty}_0\|\textbf{f}(\tilde{\textbf q}(t,\textbf q_{\delta t}))\|~dt - \int^{+\infty}_0\|\textbf{f}(\tilde{\textbf q}(t,\textbf q_0))\|~dt) \nonumber \\
&& = - \left( \int^{+\infty}_0\|\textbf{f}(\tilde{\textbf q}(t,\textbf q_{\delta t}))\|~dt - (\int^{\delta t}_0\|\textbf{f}(\tilde{\textbf q}(t,\textbf q_0))\|~dt + \int^{+\infty}_{\delta t}\|\textbf{f}(\tilde{\textbf q}(t,\textbf q_0))\|~dt) \right) \nonumber \\
&& = \int^{\delta t}_0\|\textbf{f}(\tilde{\textbf q}(t,\textbf q_0))\|~dt >0
\end{eqnarray}
Here by definition, $\textbf{f}(\tilde{\textbf q}(\delta t,\textbf q_0)) = \textbf{f}(\tilde{\textbf q}(0,\textbf q_{\delta t}))$, so for any $t>\delta t$, $\textbf{f}(\tilde{\textbf q}(t,\textbf q_0)) = \textbf{f}(\tilde{\textbf q}(0,\textbf q_{t}))$ holds.

The matrices $D$ and $Q$ are numerically constructed according to Eq.~(\ref{Eq:D}) and Eq.~(\ref{Eq:Q}). We plot the generalized landscape $\phi$ in Figure 5 (global view) and Figure 6 (local view near the saddle point, the color is re-scaled for better visibility). We plot the vector field $\textbf{f}$ of the system, the dissipative part $\textbf{f}_d$, and the conservative part $\textbf{f}_c$ in Figure 11 and Figure 12. The numeric method automatically re-explores the saddle point $G$ and locates the boundary of the attractive basins in Figure 6, whose approximate linearity suggests an interesting interchangeability between allele frequency and linkage disequilibrium (Weinreich et al., 2013).

\section{Discussion\label{sec:discussion}}
The controversies of adaptive landscape have been for different reasons, even Wright (1988) himself had been discussing this concept for more than half a century. One of the main concerns is whether adaptive landscape exists in models with non-gradient dynamics as most realistic biological models are not gradient. We tried to answer this question in a specific population model under selection and recombination, which presents a simplest non-gradient model with more than one biological forces and more than one dimensions.

We found that a part of the problem comes from the definition. Wright's original description of landscape asked for both the intuitive construction as $\overline W$ and the gradient property. As it turns out, these two aspects cannot usually be achieved in a single landscape. 
The fitness landscape can easily fail as other forces will drag a population away from the fitness peak. Even when there is only selection, a population's trajectory does not necessarily follow the gradient of the fitness landscape. 
The generalized landscape we employed were shown to exist in many non-gradient models; and as long as it exists, the Lyapunov property makes sure that a population always moves towards an adaptive peak. Nevertheless, in many cases, it is true that constructing generalized landscape explicitly is not easy.

Sometimes, the definitions of different landscapes can overlap. 
The mean fitness landscape can be a legitimate generalized landscape when there is only selection or if selection is very strong (Section \ref{Sec:Case-1}). 
The gradient landscape is a special case of the generalized landscape when the system dynamics is gradient (corresponding to $D=\textbf{1},~Q=\textbf 0$ in Eq.~(\ref{Eq:decomp})). Mean fitness landscape may or may not be a gradient landscape.
For example, the selection-only model is not gradient even in the original $p_i$ space. We consider the constraint in Eq.~(\ref{Eq:constr1}) and calculate the curl of Eq.~(\ref{Eq:4d}) with the current fitness settings:
\begin{eqnarray}
&& \nabla\times \textbf f = \bigg(\frac{s_1(1+s_1)(p_2-p_3)}{(1+s_2p_1+s_1p_2+s_1p_3)^2},~
\frac{s_2p_3-s_1p_1-s_1s_2p_1+s_1s_2p_3}{(1+s_2p_1+s_1p_2+s_1p_3)^2}~ \nonumber \\
&& \qquad ,~\frac{-s_2p_2+s_1p_1+s_1s_2p_1-s_1s_2p_2}{(1+s_2p_1+s_1p_2+s_1p_3)^2}\bigg). 
\end{eqnarray}
which is not 0 whether or not we consider the further constraint $p_2=p_3$: $\nabla\times \textbf f = (-s_2p_2+s_1p_1+s_1s_2p_1-s_1s_2p_2)/(1+s_2p_1+2s_1p_2)^2 \neq 0$. Even in this simplest model with pure selection, the mean fitness landscape is not a gradient landscape, but a generalized adaptive landscape.

Based on the generalized landscape, the system dynamics is decomposed into two parts:
\begin{eqnarray}
\text{Deterministic dynamics = dissipative part + conservative part}, 
\label{Eq:word-1} \nonumber
\end{eqnarray}
where
\begin{eqnarray}
&& \text{dissipative part = gradient part + one part of non-gradient force}, \nonumber \\
&& \text{conservative part = the other part of non-gradient force}.
\nonumber
\end{eqnarray}
Intuitively, a part of the non-gradient dynamics also contributes to the increase of adaptiveness.
The decomposition takes an energy-view of the system: When energy dissipates (decreases), stability (adaptiveness) increases, corresponding to that $D\nabla\phi$ drives the population towards the local adaptive peak. When energy conserves, stability (adaptiveness) does not change, corresponding to that $Q\nabla\phi$ drives the population on equi-adaptiveness surfaces of the landscape (neutral evolution). In the extreme case $D=0,~Q\neq0$, the dynamics is completely rotational and the uphill rate is 0 (limit cycle). 

One feature of the generalized landscape is the coordinate-independency. It allows us to construct a generalized landscape in a space with simpler dynamical presentation (for example the $p_i$ space) and use the landscape in another space (for example the $q_i$ space) with richer biological meaningfulness. The conclusion of Section \ref{Sec:CordInvar} immediately provides a way to construct a generalized landscape in any other space. 

Another feature of the generalized landscape is its non-uniqueness, as is shown by the joint Case of 1 and 2 (Section \ref{sec:construction}). This feature is consistent to the non-uniqueness of the Lyapunov function (Haddad and Chellaboina, 2008). 
It comes from the degrees of freedom in determining $\phi,~D,$ and $Q$ in Eq.~(\ref{Eq:decomp}). There are additional constraints to Eq.~(\ref{Eq:decomp}) in stochastic models, which makes the construction unique (Ao, 2005). This non-uniqueness might be interpreted as that the stability information is incomplete in deterministic models (for example, the relative stabilities of two adaptive peaks cannot be determined). This property, if not strengthening, does not weaken the existence of generalized landscape in non-gradient systems.

Although we constructed generalized landscapes in the current model, it is still quite hard to find constructions for more complex high-dimensional models in general. For example, in the bi-peaked case (Section \ref{Sec:case-3}), we were not able to find the explicit form of landscape. The heuristic construction uses the trajectory information, but the constructed landscape as a scalar function is always path-independent (Robinson, 2004). It is good enough to capture the near-linear boundary between the two attractive basins (Figure 5) but is of limited numerical accuracy overall. 

\section{Conclusion}
We constructed the generalized adaptive landscape for a two-dimensional population model under the interactions of selection and recombination. The landscape visualizes the global stable states and the evolutionary trajectories of the model without bias, based on which we analyzed the non-gradient dynamics of the model in different parameter ranges. We also proved its applicability in any coordinate space and concluded that it is a possible candidate for continuing the exploration of Wright's theory for complex dynamics. 

\section*{References}
\small
Andrews, P. W. (2002). From teratocarcinomas to embryonic stem cells. \textit{Philosophical Transactions of the Royal Society B: Biological Sciences, 357}(1420), 405-417. \\
Ao, P. (2004). Potential in stochastic differential equations: novel construction. \textit{Journal of Physics A: Mathematical and General, 37(3)}: L25-L30. \\
Ao, P. (2005). Laws in Darwinian evolutionary theory. \textit{Physics of Life Reviews, 2(2)}: 117-156. \\
Ao, P. (2008). Emerging of Stochastic Dynamical Equalities and Steady State Thermodynamics from Darwinian Dynamics. \textit{Communications in Theoretical Physics, 49(5)}: 1073-1090. \\
Ao, P. (2009). Global view of bionetwork dynamics: adaptive landscape. \textit{Journal of Genetics and Genomics, 36(2)}: 63-73. \\
Ao, P., Galas, D., Hood, L., and Zhu, X. (2008). Cancer as robust intrinsic state of endogenous molecular-cellular network shaped by evolution. \textit{Medical Hypotheses, 70(3)}: 678-684. \\
Arnold, S. J., Pfrender, M. E., and Jones, A. G. (2001). The adaptive landscape as a conceptual bridge between micro-and macroevolution. \textit{Genetica, 112(1)}: 9-32. \\
Barton, N. H. and Coe, J. B. (2009). On the application of statistical physics to evolutionary biology. \textit{Journal of Theoretical Biology, 259(2)}: 317-324. \\
Carneiro, M. and Hartl, D. L. (2010). Adaptive landscapes and protein evolution. \textit{Proceedings of the National Academy of Sciences, 107}(suppl 1), 1747-1751. \\
Delbr{\'u}ck, M. (1949). G{\'e}n{\'e}tique du bact{\'e}riophage. \textit{Unit{\'e}s Biologiques Dou{\'e}es de Continuit{\'e} G{\'e}n{\'e}tique, 8}: 91-103. \\
Dill, K. A., and Chan, H. S. (1997). From Levinthal to pathways to funnels. \textit{Nature Structural Biology, 4}(1), 10-19. \\
Edwards, A. W. F. (2000) ``Sewall Wright's Equation $\Delta q=(q (1-q)\partial w/\partial q)/2w$." \textit{Theoretical Population Biology, 57}(1): 67-70. \\
Ewens, W. J. (2004) \textit{Mathematical Population Genetics: I. Theoretical Introduction}. Berlin: Springer-Verlag. \\
Gavrilets, S. (2004). \textit{Fitness landscapes and the origin of species (MPB-41)}. Princeton, NJ: Princeton University Press. \\
Ge, H. and Qian, H. (2012). Landscapes of non-gradient dynamics without detailed balance: Stable limit cycles and multiple attractors. \textit{Chaos: An Interdisciplinary Journal of Nonlinear Science, 22(2)}: 023140. \\
Haddad, W. M. and Chellaboina, V. S. (2008). \textit{Nonlinear Dynamical Systems and Control: A Lyapunov-Based Approach}. Princeton University Press: Princeton. \\
Hopfield, J. J. (1999). Brain, neural networks, and computation. \textit{Reviews of Modern Physics, 71}(2), S431. \\
Kaplan, J. (2008). The end of the adaptive landscape metaphor? \textit{Biology $\&$ Philosophy, 23(5)}: 625-638. \\
Kirkpatrick, M. and Rousset, F. (2005). Wright meets AD: not all landscapes are adaptive. \textit{Journal of Evolutionary Biology, 18}(5), 1166-1169. \\
Kwon, C., Ao, P., and Thouless, D. J. (2005). Structure of stochastic dynamics near fixed points. \textit{Proceedings of the National Academy of Sciences, 102}(37): 13029-13033. \\
Lande, R. (1976). Natural selection and random genetic drift in phenotypic evolution. \textit{Evolution, 30}(2): 314-334. \\
Lapidus, S., Han, B., and Wang, J. (2008). Intrinsic noise, dissipation cost, and robustness of cellular networks: The underlying energy landscape of MAPK signal transduction. \textit{Proceedings of the National Academy of Sciences, 105}(16), 6039-6044. \\
Lyapunov, A. M. (1992). The general problem of the stability of motion. \textit{International Journal of Control, 55}(3): 531-534. \\
Ma, Y., Tan, Q., Yuan, R., Yuan, B., and Ao, P. (2014). Potential function in a continuous dissipative chaotic system: Decomposition scheme and role of strange attractor. \textit{International Journal of Bifurcation and Chaos, 24}(02): 1450015. \\
Qian, H. (2005). Cycle kinetics, steady state thermodynamics and motors—a paradigm for living matter physics. \textit{Journal of Physics: Condensed Matter, 17}(47): S3783-S3794. \\
Rice, S. H. (2004). \textit{Evolutionary Theory: Mathematical and Conceptual Foundations}. Sunderland: Sinauer Associates. \\
Robinson, R. C. (2004). \textit{An Introduction to Dynamical Systems: Continuous and Discrete}. New York: Prentice Hall. \\
Tang, Y., Yuan, R., and Ma, Y. (2013). Dynamical behaviors determined by the Lyapunov function in competitive Lotka-Volterra systems. \textit{Physical Review E, 87}(1): 012708. \\
Volpe, G., Helden, L., Brettschneider, T., Wehr, J., and Bechinger, C. (2010). Influence of Noise on Force Measurements. \textit{Physical Review Letters, 104}(17): 170602. \\
Waddington, C. H. (1957). \textit{The Strategy of the Genes: A Discussion of Some Aspect of Theoretical Biology.} New York: The MacMillan Company. \\
Wang, J., Huang, B., Xia, X., and Sun, Z. (2006). Funneled landscape leads to robustness of cell networks: yeast cell cycle. \textit{PLoS Computational Biology, 2}(11), e147. \\
Wang, J., Xu, L., and Wang, E. (2008). Potential landscape and flux framework of nonequilibrium networks: Robustness, dissipation, and coherence of biochemical oscillations. \textit{Proceedings of the National Academy of Sciences, 105}(34): 12271-12276. \\
Weinreich, D. M., Delaney, N. F., DePristo, M. A., and Hartl, D. L. (2006). Darwinian evolution can follow only very few mutational paths to fitter proteins. \textit{Science, 312}(5770): 111-114. \\
Weinreich, D. M., Sindi, S., and Watson, R. A. (2013). Finding the boundary between evolutionary basins of attraction, and implications for Wright's fitness landscape analogy. \textit{Journal of Statistical Mechanics: Theory and Experiment, 2013}(01): P01001. \\
Wright, S. (1932). The roles of mutation, inbreeding, crossbreeding, and selection in evolution. \textit{Proceedings of the Sixth International Congress of Genetics, 1}: 356-366. \\
Wright, S. (1988). Surfaces of selective value revisited. \textit{American Naturalist, 131}(1): 115-123. \\
Xu, S., Jiao, S. Y., Jiang, P. Y., and Ao, P. (2014). Two-time-scale population evolution on a singular landscape. \textit{Physical Review E, 89}(1): 012724. \\
Yuan, R., Ma, Y., Yuan, B., and Ao, P. (2013). Lyapunov function as potential function: A dynamical equivalence. \textit{Chinese Physics B, 23}(1): 010505. \\
\textit{Physical Review E, 87}(6): 062109. \\
Zhu, X., Yin, L., Hood, L., and Ao, P. (2004). Calculating biological behaviors of epigenetic states in the phage λ life cycle. \textit{Functional $\&$ Integrative Genomics, 4}(3): 188-195.

\newpage

\section*{Appendix}

\appendix
\setcounter{equation}{0}
\renewcommand{\theequation}{A-\arabic{equation}}

\section{Dynamical factors in Case 1}
\label{Sec:GQ}

We calculate the expressions of $D_{11}$ and $Q_{21}$ as:
\begin{eqnarray}
D_{11} = &&\Bigg(\frac{2 (s_1 - 2 q_1 s_1 + q_1 s_2) \Big(q_2 \Big((-1 + 2 q_1) s_1 + s_2 - q_1 s_2\Big) + (1 - q_1) q_1 (s_1 - 2 q_1 s_1 + q_1 s_2)\Big)}{1 + 2 q_1 s_1 + (q_2 + q_1^2) (-2 s_1 + s_2)} + \nonumber \\
&& ~~(-2 s_1 + s_2)\Bigg(-q_2 +
\frac{-\Big(q_2 + (-1 + q_1) q_1\Big)^2 (2 s_1 + s_1^2 - s_2) + q_2 (1 + s_2)}{\Big(1 + 2 q_1 s_1 + (q_2 + q_1^2) (-2 s_1 + s_2)\Big)^2} - \nonumber \\
&& ~~\frac{r \Big(-\Big(q_2 + (-1 + q_1) q_1 \Big )^2 (2 s_1 + s_1^2 - s_2) + q_2 (1 + s_2)\Big)}{(1 + 2 q_1 s_1 + (q_2 + q_1^2) (-2 s_1 + s_2))^2}\Bigg) \nonumber \\
&&\Bigg) \Bigg / \Big((-2 s_1 + s_2)^2 + 4 (s_1 - 2 q_1 s_1 + q_1 s_2)^2\Big).
\end{eqnarray}
and
\begin{eqnarray}
Q_{21} = && \Bigg(-\frac{(-2 s_1 + s_2) \Big(q_2 \Big((-1 + 2 q_1) s_1 + s_2 - q_1 s_2\Big) + (1 - q_1) q_1 (s_1 - 2 q_1 s_1 + q_1 s_2)\Big)}{1 + 2 q_1 s_1 + (q_2 + q_1^2) (-2 s_1 + s_2)} + \nonumber \\
&& ~~2 (s_1 - 2 q_1 s_1 + q_1 s_2) \Bigg(-q_2 + \frac{-\Big(q_2 + (-1 + q_1) q_1\Big)^2 (2 s_1 + s_1^2 - s_2) + q_2 (1 + s_2)}{\Big(1 + 2 q_1 s_1 + (q_2 + q_1^2) (-2 s_1 + s_2)\Big)^2} - \nonumber \\
&& ~~\frac{r \Big(-\Big(q_2 + (-1 + q_1) q_1\Big)^2 (2 s_1 + s_1^2 - s_2) + q_2 (1 + s_2)\Big)}{\Big(1 + 2 q_1 s_1 + (q_2 + q_1^2) (-2 s_1 + s_2)\Big)^2}\Bigg) \nonumber \\
&& \Bigg) \Bigg/ \Big((-2 s_1 + s_2)^2 + 4 (s_1 - 2 q_1 s_1 + q_1 s_2)^2\Big).
\end{eqnarray}

\section{Continuity of the numerical construction in the bi-peaked case $-1\leq s_1<0<s_2$ and $r\neq 0$}
\label{BiNumr}

Consider a point $\textbf{q}_0$ in the attractive basin of $E$. The specific trajectory $\tilde{\textbf{q}}(t,\textbf{q})$
starting at $\textbf{q}$ in the neighborhood $\mathcal{O}_0$ of $\textbf{q}_0$
will all converge to the stable fixed point $E$
according to the $Poincar\acute{e}-Bendixson$ theorem (Hirsh and Smale, 1974).

For any given $\epsilon>0$, select a neighborhood $\mathcal{O}_A$ of A
such that $\tilde{\textbf{q}}(t+T,\textbf{q}_0)$ will be in $\mathcal{O}_A$ for all
$\textbf{q}\in\mathcal{O}_0$, $t>0$, and a fixed $T>0$.
Inside the neighborhood $\mathcal{O}_A$,
there are constants $c,~K>0$ satisfying (Robinson, 2004)
\begin{eqnarray}
\|\tilde{\textbf q}(t+T,\textbf{q}_0)\|\leq K\|\tilde{\textbf q}(T,\textbf{q}_0)\|e^{-ct}.
\end{eqnarray}
Inside $\mathcal{O}_A$, the integration is dominated by the linearized part $D\textbf{f}_A$
at $(0,0)$ which is assumed to be non-degenerated,
and there is a constant $c'>0$ that
\begin{eqnarray}
&&\int^{+\infty}_T\|\textbf{f}(\tilde{\textbf{q}}(t,\textbf{q}_0))\|~dt 
\leq\int^{+\infty}_Tc'\|D\textbf{f}_A\cdot\tilde{\textbf{q}}(t,\textbf{q}_0)\|~dt \nonumber\\
&&\leq\int^{+\infty}_Tc'\|D\textbf{f}_A\|\cdot\|\tilde{\textbf{q}}(t,\textbf{q}_0)\|~dt 
\leq\int^{+\infty}_0c'\|D\textbf{f}_A\|\cdot K\|\tilde{\textbf q}(T,\textbf{q}_0)\|e^{-ct}~dt \nonumber\\
&&= c'K\cdot \|D\textbf{f}_A\|\cdot \|\tilde{\textbf q}(T,\textbf{q}_0)\|\int^{+\infty}_0e^{-ct}~dt \nonumber \\
&& = \frac{c'K\cdot \|D\textbf{f}_A\|}{c}\cdot \|\tilde{\textbf q}(T,\textbf{q}_0)\|. 
\end{eqnarray}
Choose the neighborhood $\mathcal{O}_A$ with radius smaller than $\dfrac{c}{4c'K\cdot \|Df_A\|}\epsilon>0$.
Because the flow is continuous with respect to $t\times\textbf{q}$,
the neighborhood $\mathcal{O}_0$ can be also selected small enough such that
\begin{eqnarray}
\left|\int^{T}_0\|\textbf{f}(\tilde{\textbf{q}}(t,\textbf{q}_1))\|~dt-\int^{T}_0\|\textbf{f}(\tilde{\textbf{q}}(t,\textbf{q}_2))\|~dt\right|\leq\frac{\epsilon}{2}.
\end{eqnarray}
where $\textbf{q}_1,~\textbf{q}_2\in\mathcal{O}_0$.
In all, $|L(\textbf{q}_0)-L(\textbf{q})|<\epsilon$ for all $\textbf{q}\in\mathcal{O}_0$, therefore $L$ is continuous at $\textbf{q}_0$.

Near the boundary between the two attractive basins, which is the stable manifold of the saddle $G$,
the orbit will first follow the stable manifold to the neighborhood of the saddle, then to either $E$
or B following the unstable manifold of the saddle. There is a constant gap of the geometry length
from the left side and the right sides. This gap has been filled in Eq.~(\ref{Eq:geo construction})
by subtracting the saddle-to-stable length, either $L_E$ or $L_F$, depending on whether it is on the left or right side.

\section*{Figures}
\begin{figure}[h]
\centering
\includegraphics[scale=0.35]{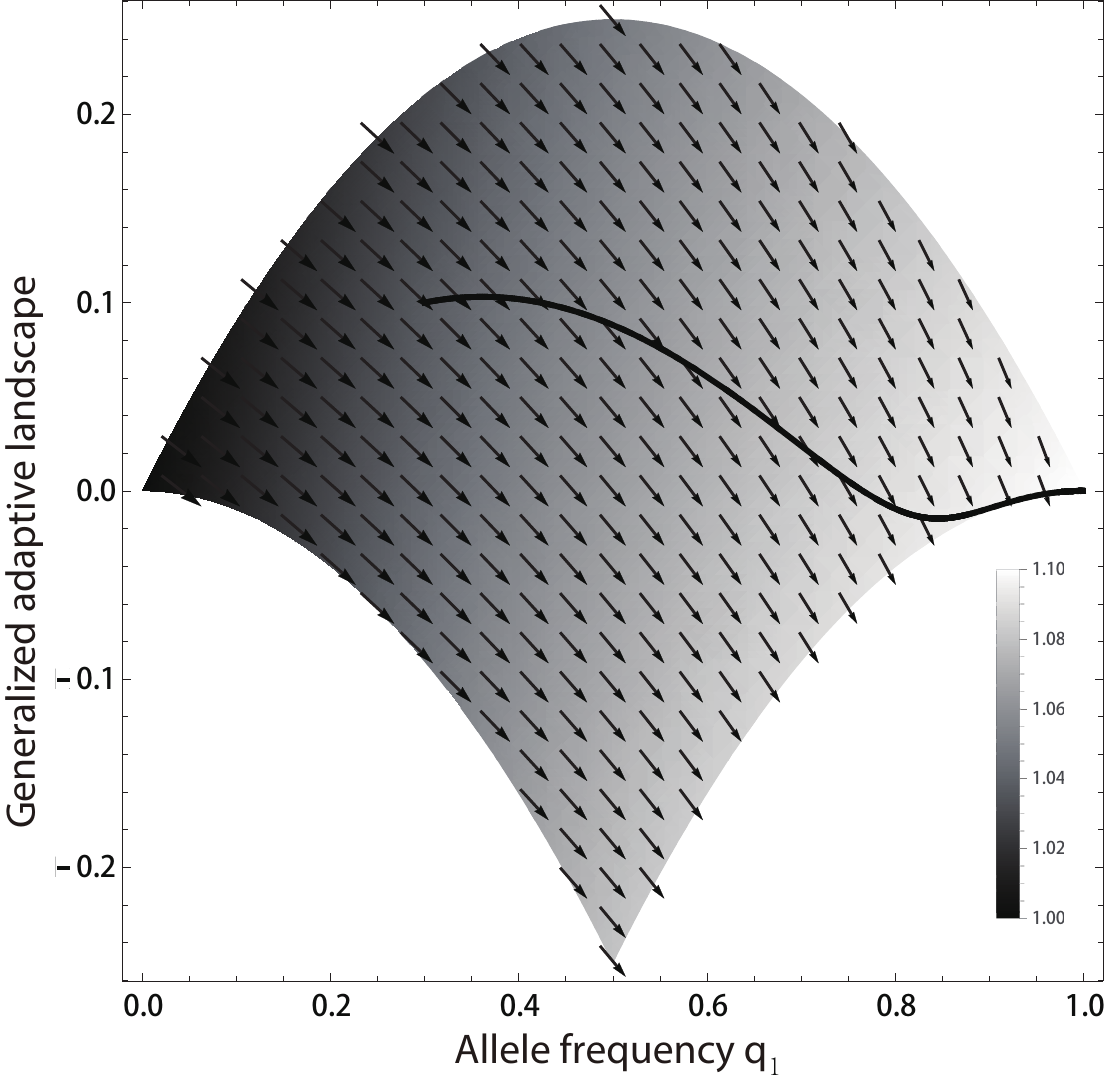}
\caption{A typical generalized landscape in the $q_1\times q_2$ space. The white part denotes states with higher adaptiveness and the black part denotes lower adaptiveness. The arrows denote the gradient field of the landscape. The bold line denotes an evolutionary trajectory towards the global stable state (0,~1). The trajectory does not generally follow the gradient direction.}
\end{figure}

\begin{figure}[h]
\centering
\includegraphics[scale=0.5]{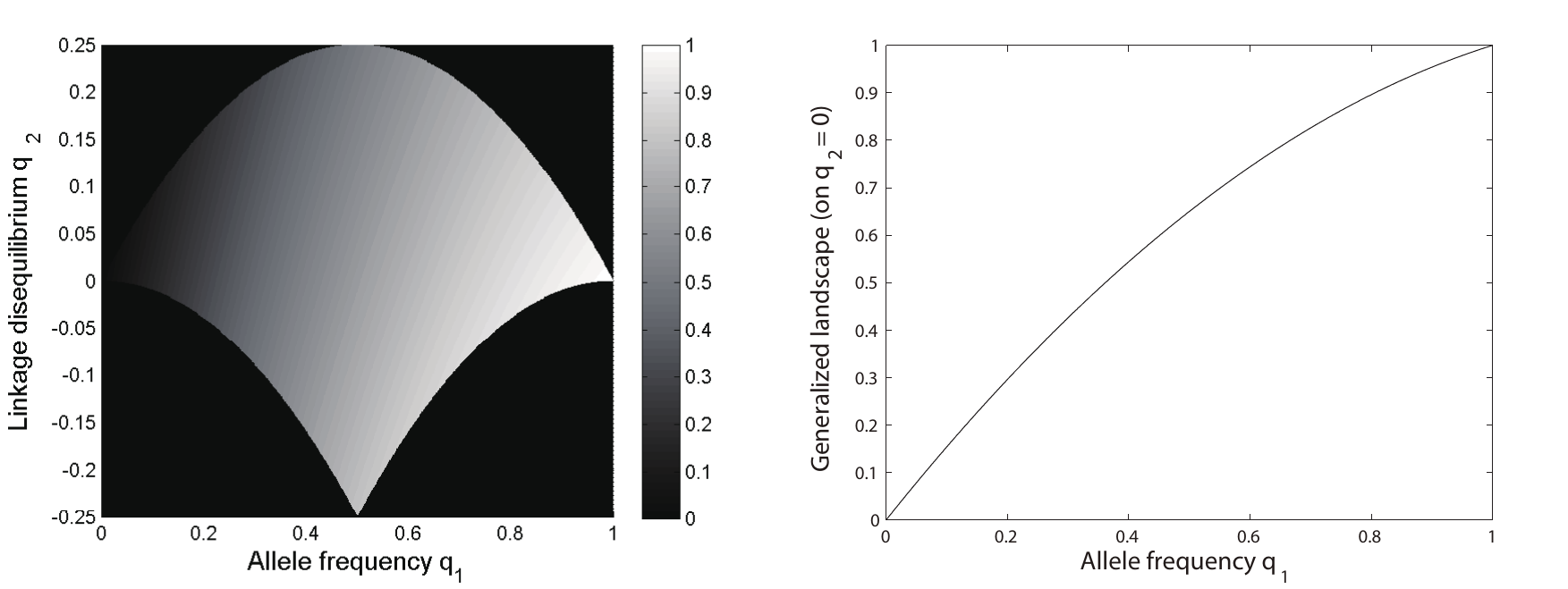}
\caption{Generalized landscape $\phi=\overline W$ and its projection on the line $q_2=0$ in Case 1 with $s_1=0.08,~s_2=0.1,~r=0$.}
\end{figure}

\begin{figure}[h]
\centering
\includegraphics[scale=0.5]{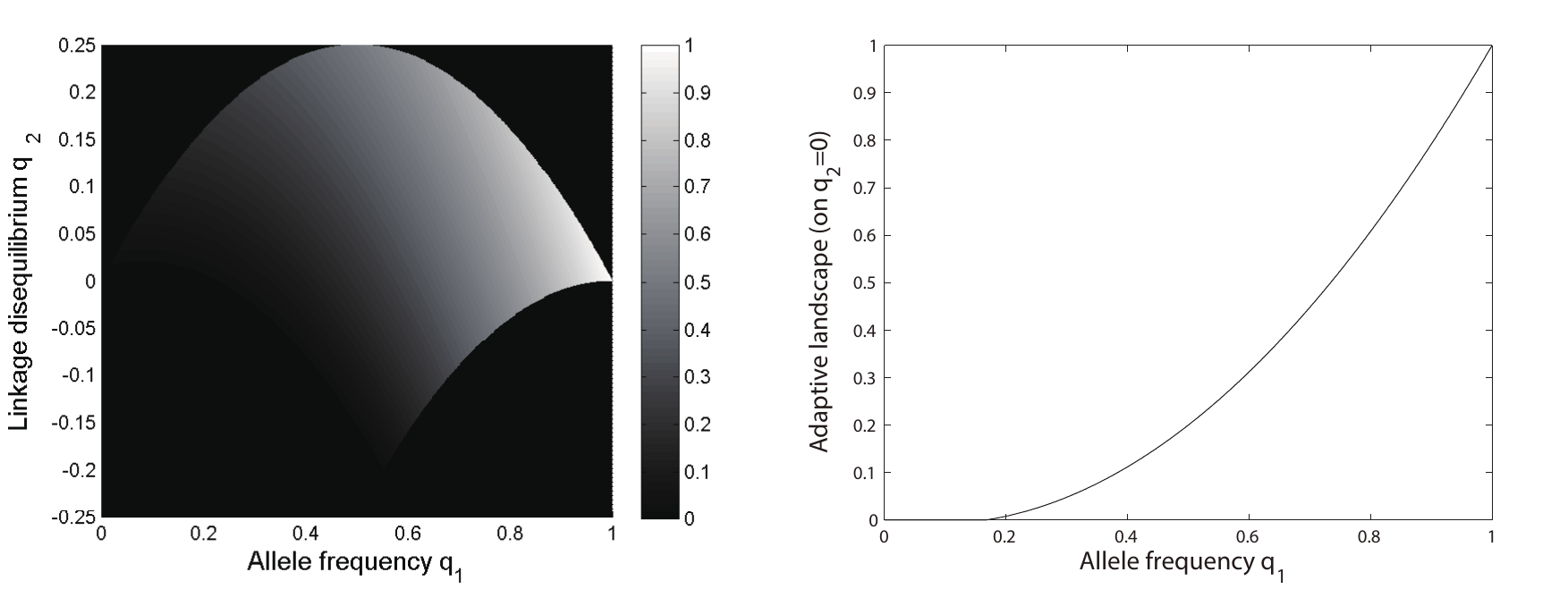}
\caption{Generalized landscape $\phi=\overline W$ and its projection on the line $q_2=0$ in Case 1 with $s_1=-0.01,~s_2=0.1,~r=0$.}
\end{figure}

\begin{figure}[h]
\centering
\includegraphics[scale=0.5]{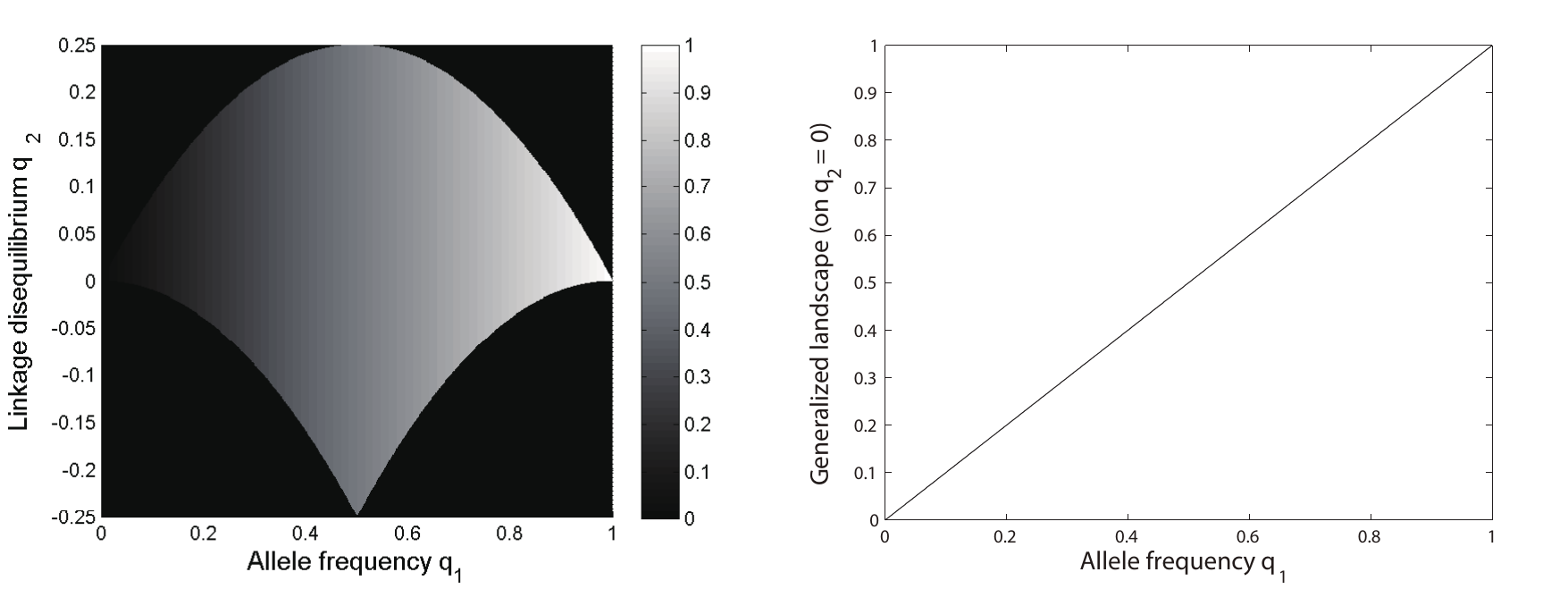}
\caption{Generalized landscape $\phi=q_1$ and its projection on the line $q_2=0$ in Case 2 with $s_1=0.08,~s_2=0.1,~r=0.2$.}
\end{figure}

\begin{figure}[h]
\centering
\includegraphics[scale=0.45]{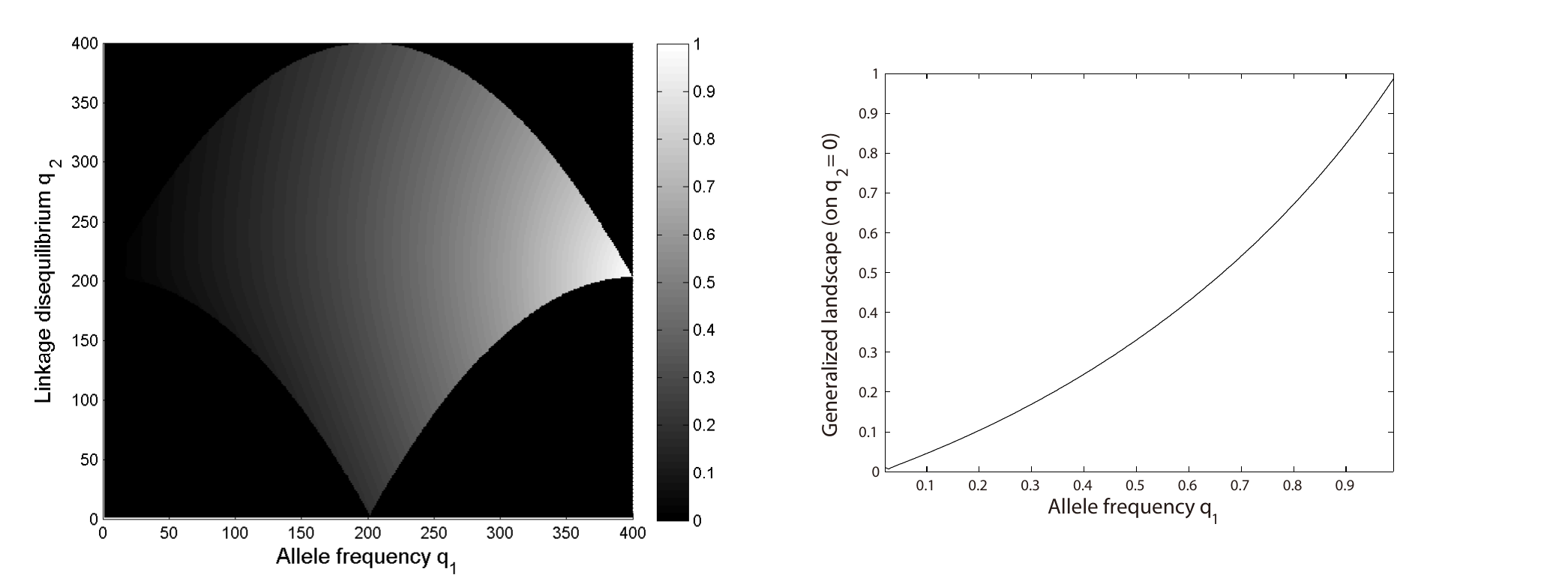}
\caption{Global view of the numerically constructed generalized landscape (without explicit form) and its projection on the line $q_2=0$ in Case 3 with $s_1=-0.01,~s_2=0.1,~r=0.125$.}
\end{figure}

\begin{figure}[h]
\centering
\includegraphics[scale=0.45]{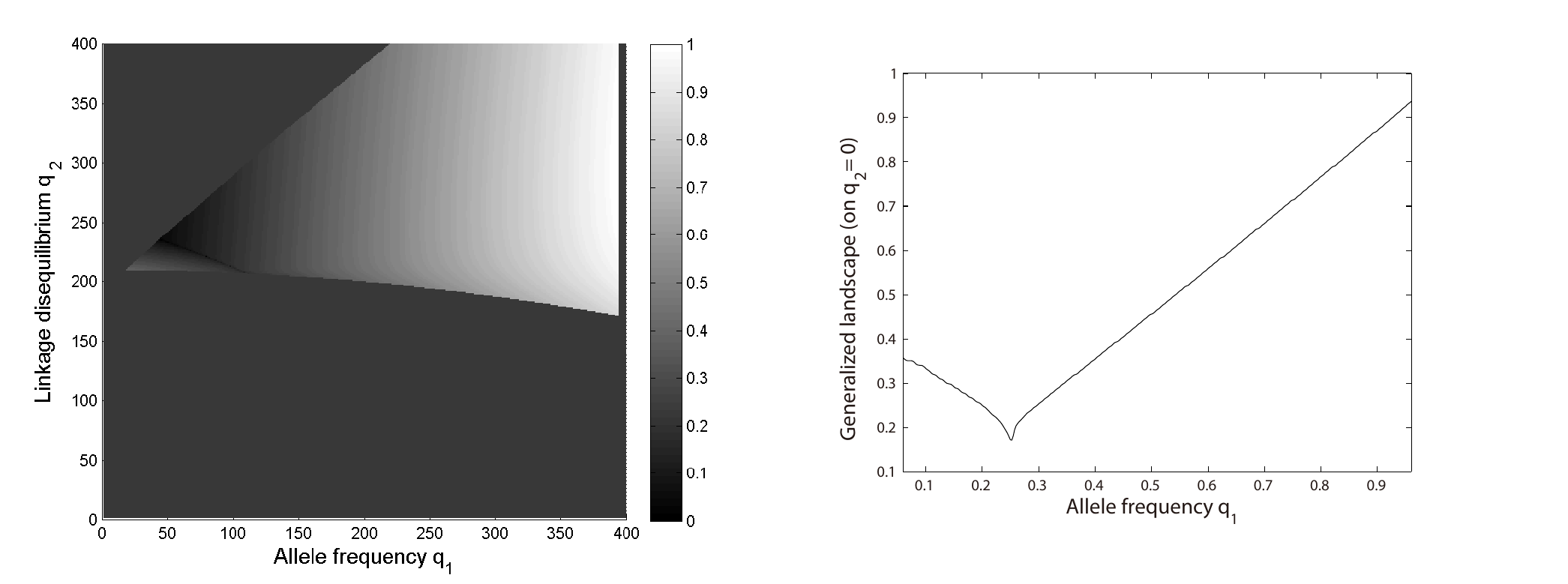}
\caption{Local view (grayscale rescaled) of the numerically constructed generalized landscape (without explicit form) and its projection on the line $q_2=0$ in Case 3 with $s_1=-0.01,~s_2=0.1,~r=0.125$.}
\end{figure}

\begin{figure}[h]
\centering
\includegraphics[scale=0.5]{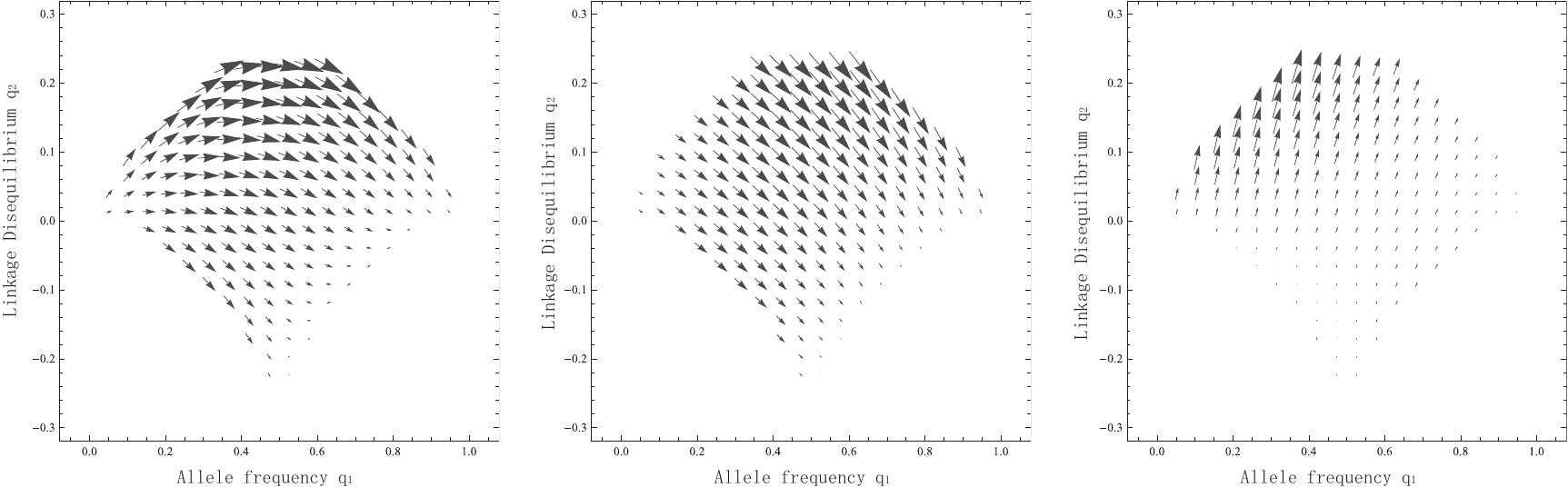}
\caption{Decomposition of deterministic dynamics $\textbf{f}$ (subfigure 1) into the dissipative part $\textbf{f}_d$ (subfigure 2) and the conservative part $\textbf{f}_c$ (subfigure 3) in Case 1 when $\phi=\overline W$ with $s_1=0.08,~s_2=0.1,~r=0$.}
\end{figure}

\begin{figure}[h]
\centering
\includegraphics[scale=0.5]{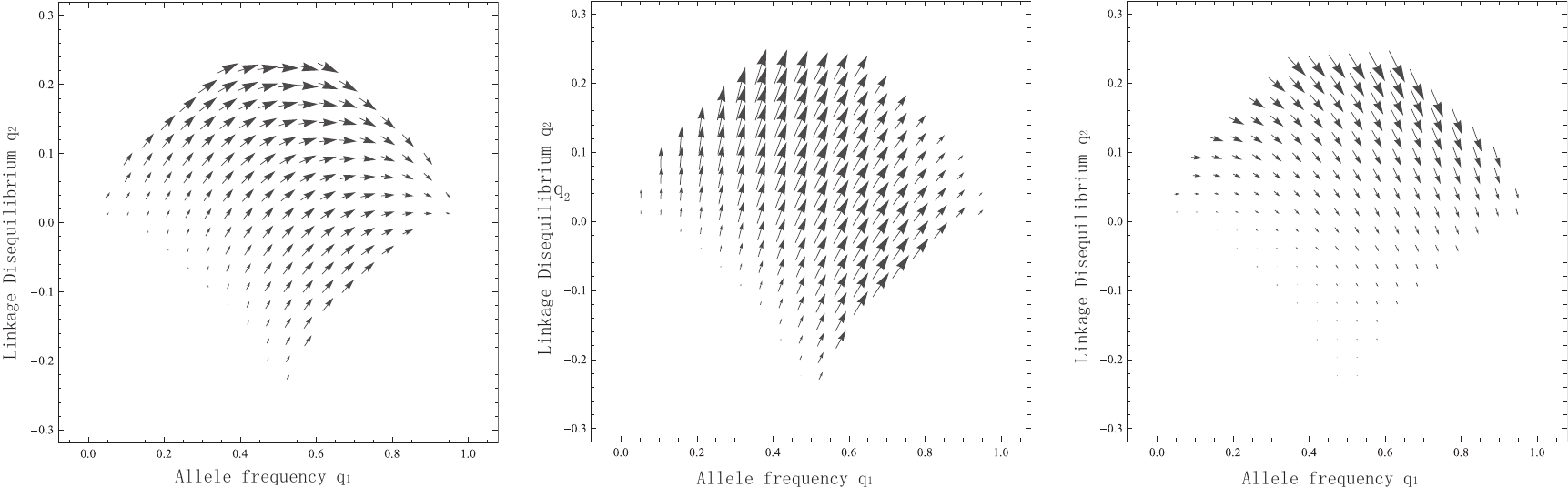}
\caption{Decomposition of deterministic dynamics $\textbf{f}$ (subfigure 1) into the dissipative part $\textbf{f}_d$ (subfigure 2) and the conservative part $\textbf{f}_c$ (subfigure 3) in Case 1 when $\phi=\overline W$ with $s_1=-0.01,~s_2=0.1,~r=0$.}
\end{figure}

\begin{figure}[h]
\centering
\includegraphics[scale=0.5]{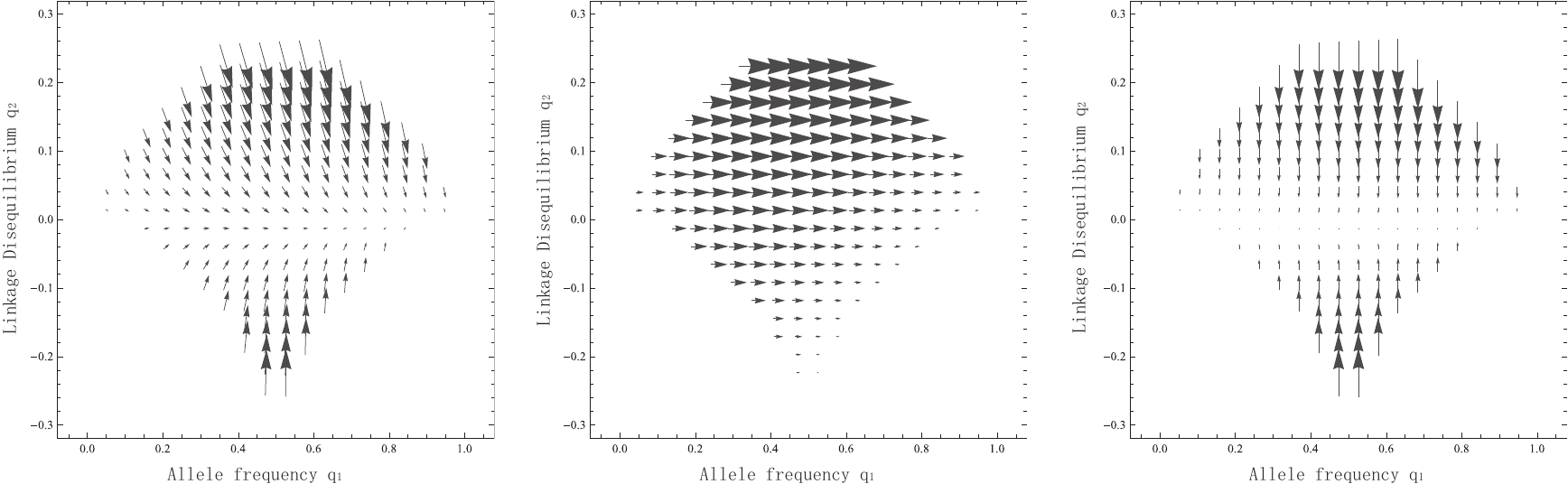}
\caption{Decomposition of deterministic dynamics $\textbf{f}$ (subfigure 1) into the dissipative part $\textbf{f}_d$ (subfigure 2) and the conservative part $\textbf{f}_c$ (subfigure 3) in Case 2 when $\phi=q_1$ with $s_1=0.08,~s_2=0.1,~r=0.2$.}
\end{figure}

\begin{figure}[h]
\centering
\includegraphics[scale=0.5]{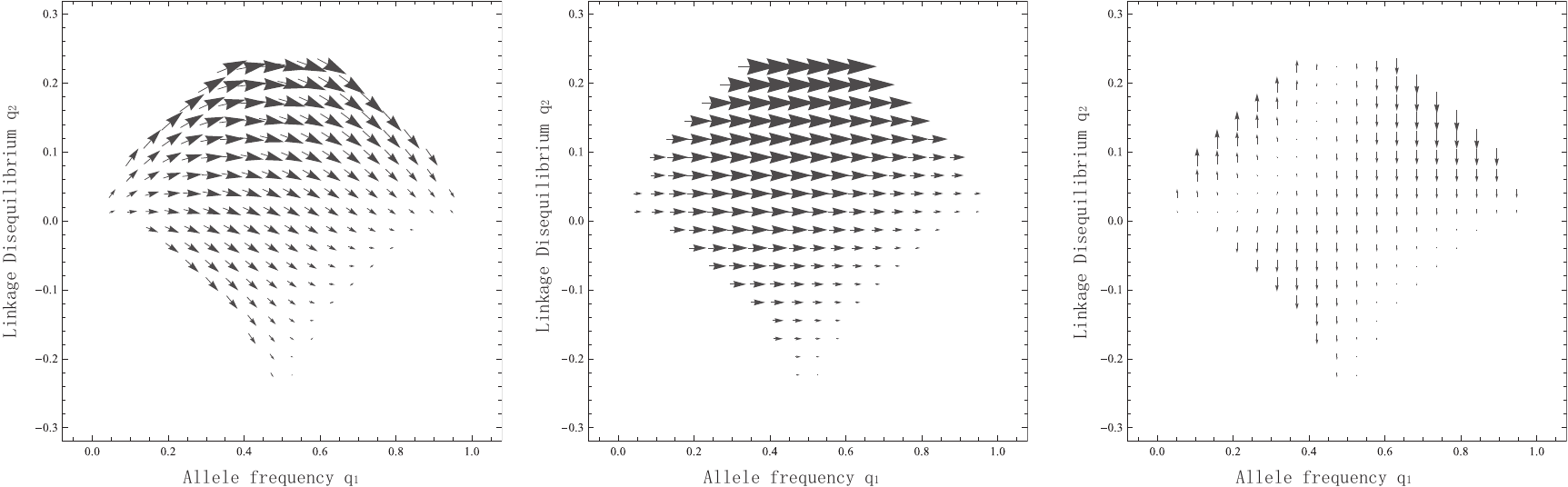}
\caption{Decomposition of deterministic dynamics $\textbf{f}$ (subfigure 1) into the dissipative part $\textbf{f}_d$ (subfigure 2) and the conservative part $\textbf{f}_c$ (subfigure 3) in Case 2 when $\phi=q_1$ with $s_1=0.08,~s_2=0.1,~r=0$.}
\end{figure}

\begin{figure}[h]
\centering
\includegraphics[scale=0.5]{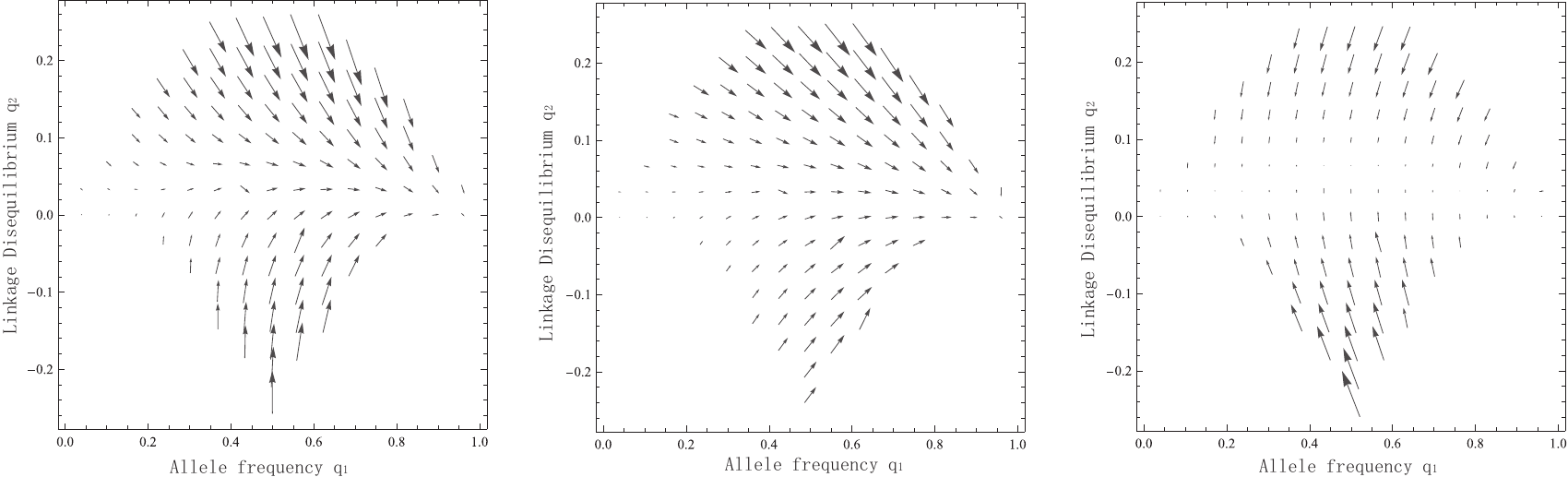}
\caption{Decomposition (global view) of deterministic dynamics $\textbf{f}$ (subfigure 1) into the dissipative part $\textbf{f}_d$ (subfigure 2) and the conservative part $\textbf{f}_c$ (subfigure 3) in Case 3 with $s_1=-0.01,~s_2=0.1,~r=0.125$.}
\end{figure}

\begin{figure}[h]
\centering
\includegraphics[scale=0.5]{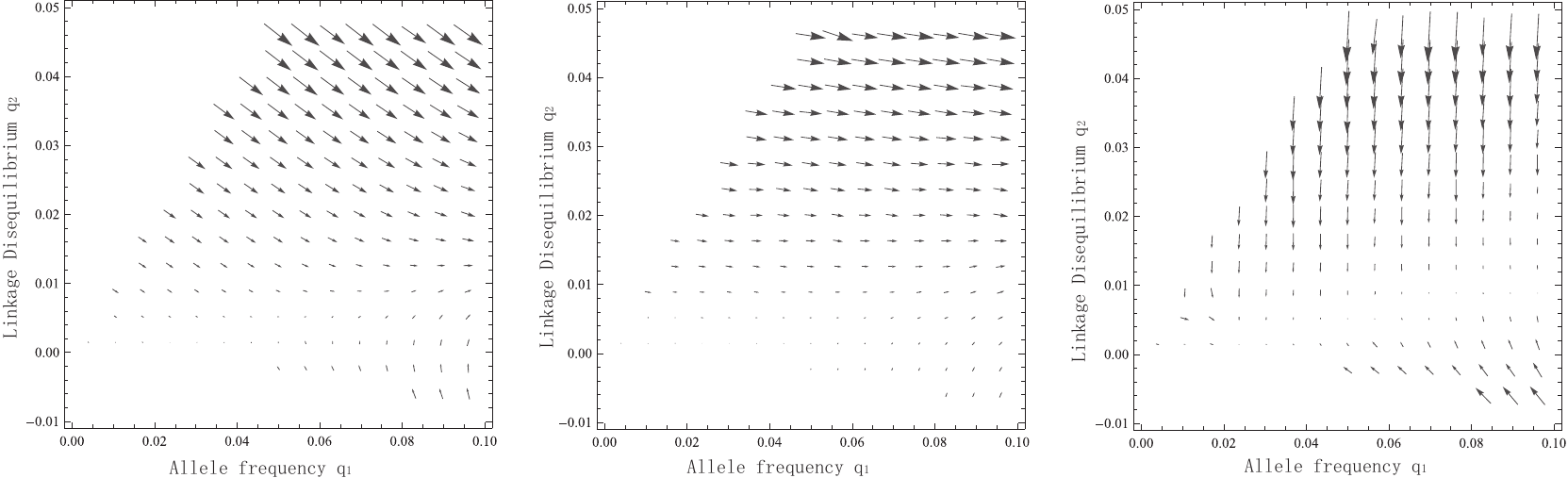}
\caption{Decomposition (local view) of deterministic dynamics $\textbf{f}$ (subfigure 1) into the dissipative part $\textbf{f}_d$ (subfigure 2) and the conservative part $\textbf{f}_c$ (subfigure 3) in Case 3 with $s_1=-0.01,~s_2=0.1,~r=0.125$.}
\end{figure}

\end{document}